\documentclass[aoas,nameyear,dvips]{arximspdf}
\usepackage{graphics}
\doi{10.1214/09-AOAS273} \volume{3} \issue{4} \pubyear{2009}
\firstpage{1382} \lastpage{1402}

\begin{document}
\begin{frontmatter}

\title{Functional data analytic approach of modeling ECG T-wave shape
to measure cardiovascular behavior\thanksref{T1}}
\thankstext{T1}{Supported in part by NISS and by Merck and Co.}
\runtitle{FDA modeling of ECG T-wave shape}

\begin{aug}
\author[A]{\fnms{Yingchun} \snm{Zhou}\ead[label=e1]{yczhou@stat.ecnu.edu.cn}\corref{}}
\and
\author[B]{\fnms{Nell} \snm{Sedransk}\ead[label=e2]{sedransk@niss.org}}
\runauthor{Y. Zhou and N. Sedransk}
\affiliation{East China Normal University and National Institute of
Statistical Sciences}
\address[A]{School of Finance and Statistics\\
East China Normal University\\
500 Dongchuan Road\\
Shanghai, 200241\\ China\\
\printead{e1}} %adresu isvedimo komanda gale!
\address[B]{19 T.W.Alexander Drive\\
Research Triangle Park\\
Durham, North Carolina 27709\\USA\\
\printead{e2}}
\end{aug}

% HISTORY:
\received{\smonth{2} \syear{2009}}
\revised{\smonth{7} \syear{2009}}

% ABSTRACT
%
\begin{abstract}
The T-wave of an electrocardiogram (ECG) represents the ventricular
repolarization that is critical in restoration of the heart muscle to a
pre-contractile
state prior to the next beat. Alterations in the T-wave reflect various
cardiac conditions;
and links between abnormal (prolonged) ventricular repolarization and
malignant arrhythmias
have been documented. Cardiac safety testing prior to approval of any
new drug currently relies
on two points of the ECG waveform: onset of the Q-wave and termination
of the T-wave; and only a
few beats are measured. Using functional data analysis, a statistical
approach extracts a common
shape for each subject (reference curve) from a sequence of beats, and
then models the deviation
of each curve in the sequence from that reference curve as a
four-dimensional vector. The representation
can be used to distinguish differences between beats or to model shape
changes in a subject's T-wave over time.
This model provides physically interpretable parameters characterizing
T-wave shape, and is robust to the
determination of the endpoint of the T-wave. Thus, this dimension
reduction methodology offers the strong
potential for definition of more robust and more informative biomarkers
of cardiac abnormalities than the QT
(or QT corrected) interval in current use.
\end{abstract}

% KEYWORDS
%

\end{frontmatter}

%s1 ###
\section{Introduction}
\begin{figure}

\includegraphics{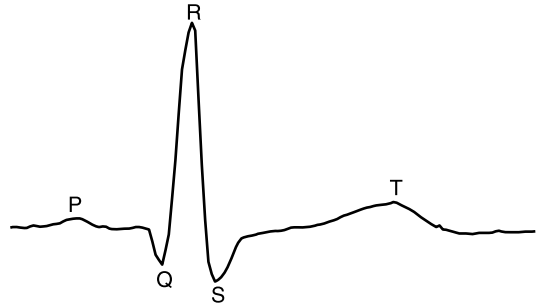}

\caption{Features of a normal ECG.}\label{fig:ECG}\vspace*{-5pt}
\end{figure}

Electrocardiograms (ECGs) are widely used to screen and monitor the
cardiac function of patients; and the behavior of the ECG wave form is
a basis for diagnosis of specific abnormalities. Important wave forms
of an ECG are marked by P, Q, R, S, T, as illustrated in Figure \ref
{fig:ECG}; these represent the changes in electrical potential as the
heart contracts and relaxes. The T-wave represents the repolarization
(or post-contractile phase) of the ventricles; and it is generally the
most labile wave in the ECG. Abnormalities in the T-wave may be
physiologic or may be externally induced, for example, by cardio-active drugs.

%f1 ###

The link between cardiac repolarization abnormalities and malignant
arrhythmias, especially torsades de pointes (TdP), which may
degenerate into ventricular fibrillation leading to sudden death, is
well documented. Since some drugs, for example, haloperidol
[\citet{FDAwebsite}] and terfenadine
[\citet{morganrothbrowncritz1993}], have been shown to cause
repolarization abnormalities, testing for cardiac safety is required
as a part of every new drug application to the \textit{FDA}; and the
\textit{FDA} places very stringent constraints on the allowable
prolongation of the repolarization process.

%f2 ###
\begin{figure}[b]

\includegraphics{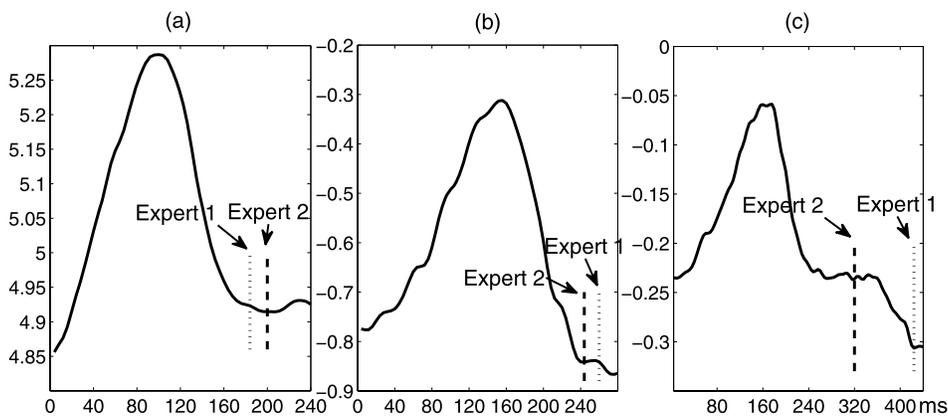}

\caption{Two cardiologists' marks of
T-wave ends for 3 beats in a QT Dataset. The differences are 17
milliseconds \textup{(a)}, 15 milliseconds \textup{(b)} and 104 milliseconds \textup{(c)}.}\label{fig:phydiff103117114}\vspace*{-5pt}
\end{figure}

The current measure for cardiac safety that is used in drug
development and drug approval is prolongation of the QT interval.
The premise for using this measure is the evidence that the
particular repolarization aberration, TdP, is either preceded by or
characterized by a pro-arrhythmia defined in terms of delayed
termination of the T-wave; more recently increased heterogeneity of
T-waves has also been implicated [\citet{couderc2009}]. The QT
interval was first put forward in the 1920s and has been in
continual use since, with little modification and with cardiologists
personally marking the two critical points on the ECG: the
initiation of the Q-wave and the termination of the T-wave. From a
practical point of view, in a normal ECG of good quality, there is
relatively little difficulty in the determination of the onset of
the Q-wave even though an ECG has no actual ``baseline.'' However,
measurement of the QT interval also relies on accurate and
reproducible determination of the endpoint of the T-wave, which is a
greater challenge, as can be seen from Figure \ref{fig:phydiff103117114}. The differences
between two cardiologists' marks are 17 milliseconds in (a), 15
milliseconds in (b) and 104 milliseconds in (c). For QT analyses
presented to the \textit{FDA}, exceedance of 10 milliseconds for the
maximal time difference between drug and control over all time
points for a single patient calls into question the cardiac safety,
requiring further discussion, at the least, before considering
approval of the drug.

Thus, weaknesses of the current measurement method are four-fold: (1)
QT as a cardiac safety indicator is predicated on detecting a
heart-rhythm change associated (but not exclusively) with a
particular cardiac arrhythmia. Slow trends, abrupt shifts in T-wave
morphology, early precursors for T-wave changes and/or episodic
events are not necessarily reflected. (2) The measurement ignores
the information in the shape of the curve (the T-wave morphology),
relying instead upon only two points of the complex curve. (3)
Accuracy and the reproducibility of the measurement based on the
endpoint of the T-wave depend on the sharpness of the T-wave form.
(4) Current measurement practice calls for measuring and then
averaging a sequence of a few (usually three) ``highly similar''
beats, most often from a 10-second record. (This 10-second sequence
is often preselected by an algorithm that excludes difficult-to-read
``outlier'' beats and then captures a sequence for a near-constant
heart rate, following 90 seconds of minuscule heart-rate variation.
In this case, the cardiologist does not see the complete ECG, but
only the selected beat sequence.)

In this paper a statistical model of the T-wave shape is
constructed based on function data analysis (FDA), using all the data
in an extended (minutes or longer) ECG record. This model has four
interpretable parameters and performs well in describing both normal
and arrhythmia ECGs available in public libraries. Because the
parametrization of the model accounts for the morphology of the
entire T-wave, it is particularly useful for describing changes in
the repolarization process. It also has the significant advantage
that it is robust to the determination of the onset or the end of
the T-wave.

Effectively, this functional data analysis
approach decomposes a sequence of T-waves for an individual subject
into a reference curve (representing the common shape of the
T-waves) and a four-dimensional representation of the deviation of
each individual T-wave in the sequence. Inference about changes in
cardiac function within the sequence can now be analyzed through the
four-dimensional representation of individual T-waves. For multiple
ECG sequences for the same subject, the reference curves for the
individual sequences can be treated as data to again be analyzed by
the appropriate construction of a (superpopulation) hyper-reference
curve (representing the common shape of the reference curves) and
four-dimensional representations of the ``hyper-deviations'' of each
reference curve from the hyper-curve. In this fashion, the
hyper-deviations can be used to analyze longer scale processes such
as diurnal effects or shifts in baseline ECGs for long-term
experiments.

Alternative methods to model the shape of ECG waves including the
T-wave are principal component analysis (PCA)
[\citet{lagunamoodygarciagoldbergermark1999}] and Gaussian models
[\citet{clifford2006}]. Both methods fit the wave forms quite well
and reduce the dimension of the data significantly. However, in
terms of interpretation of model parameters, neither of them does
well. The principal components and loadings in PCA do not provide
physical interpretation of ECGs. The location and scale parameters
in a Gaussian model may reveal some information about the shape of a
T-wave, but they are not robust to small changes in T-waves such as
those caused by noise. These features are shown in specific examples
in Section~\ref{sec4}.

The major difference between this FDA model
and other models is that this model has a common reference curve for
all the beats in the sequence and measures the deviation of each
wave from the common curve. So it is most useful when one wants to
compare the wave shape of a sequence of beats. All the other methods
treat each wave separately, making it harder to compare model
parameters across beats.

The ECGs examined in this research were all taken under normal clinical
conditions and are digitized; they are in the public libraries through
physionet (\href{http://www.physiotnet.org}{www.physiotnet.org}). These ECGs include both normal subjects
and subjects with various classes of arrhythmias and other cardiac
function abnormalities; there are no data available (there or
elsewhere) from actual QT studies since this data is proprietary and is
held securely by the pharmaceutical companies.

Section \ref{sec2} describes the preprocessing of a sample of the digitized ECG
data before feeding it to the model, and lays out a general data
structure to be studied. Section \ref{sec3} describes the basic model,
illustrates model robustness to the marking of T-wave boundaries, and
shows the relation of model parameters to QT. In Section~\ref{sec4} statistical
inference is shown on how to apply the model to T-wave analysis.
Further issues and potential extension regarding the model and its
applications are discussed on Section \ref{sec5}. Section \ref{sec6} summarizes the
conclusions from this research.

\eject

%s2 ###
\section{Description of data}\label{sec2}

%s2.1 ###
\subsection{One sample of ECG series}

ECG data consist of a series of digitized waveforms, where the
digitized value is the intensity of electrical potential (in
millivolts) usually taken at a rate of 250 Hz (1 point per 4
milliseconds) or 1000~Hz (1 point per millisecond).

In order to make full use of the complete ECG record and to understand
the natural variation of the T-wave over time, a natural representation
is given by ``stacking'' aligned ECG segments of consecutive beats within
a sample. To study the morphology of the T-wave, the stacked beats are
left- and right-truncated uniformly, retaining the entire T-wave to
form an $I\times J$ matrix $X=[x_1,\ldots,x_I]^T$, where
$x_i=[x_{it_1},\ldots,x_{it_J}]$ is the digitized value of T-wave
during the interval of the $i$th beat and $I$ is the number of total
beats in this record. Beats are aligned according to their QRS
complexes since this complex is the most remarkable feature of ECG, and
it allows the easiest and least variable alignment. While there are
various methods for choosing the beginning and the end of T-waves, for
example, the threshold method and the slope method [\citet
{panickerkarnad2006}], for the purpose of this study, the beginning and
end points of the stacked T-waves are chosen to be identical for all
the T-waves and to visually capture the shape of the T-waves. As will
be shown later, the method described in this paper is robust to the
choice of the beginning and end points. Having the beginning and end
points equal for all beats is convenient for data analysis.

%s2.2 ###
\subsection{General data structure}\label{sec2.2}

A general data structure includes three levels: beat, sample and
subject, one nested within another. In a typical study there are $P$
subjects, each subject has $Q_p$ samples, and each sample has
$I_{pq}$ beats. Each beat has $J_{pq}$ time points within the T-wave
interval. The data structure and notation is as follows:
\begin{itemize}[$\bullet$]
\item Subject $p$, $p=1,\ldots,P$.
\item Sample $q$ (nested within subject $p$), $q=1,\ldots,Q_{p}$.
\item Beat $i$ (nested within subject $p$ and sample $q$), $i=1,\ldots,I_{pq}$.
\item Time point $t_j$, $j=1,\ldots,J_{pq}$.
\item Digitized value $x_{it_j}$, $i=1,\ldots,I_{pq}$, $j=1,\ldots,J_{pq}$.
\end{itemize}

Note that all the data we use in this paper are digitized ECG data
from PhysioNet (\href{http://www.physionet.org}{www.physionet.org}), a public research resource
website for physiologic signals. The sampling frequency is 250 Hz
(250 points per second).
\eject
%s3 ###
\section{T-wave modeling using functional data analysis}\label{sec3}

%s3.1 ###
\subsection{A model based on modes of variation of functional data}

Consider the data matrix $X$ for a sample ECG. Although $X$ involves
only discrete values, it reflects smooth curves of T-waves that
generate these values. As explained by Ramsay and Silverman
[\citet{ramsaysilverman1997}], this data matrix can be viewed as
functional data since each row is a function of the time points at
which they are measured, and these functions together form a family
of functions. The goal here is to characterize each function in this
family and to measure their variation.

One way to analyze
functional data is to decompose variations along nonlinear
directions from a common shape, denoted as the reference curve in this
paper. Nonlinear decomposition of curves or multivariate data has\break
received much attention in the past twenty years. Hastie and
Stuetzle\break [\citet{hastiestuetzle1989}] did pioneering work in
defining and computing principal curves, which extend the linear PCA
decomposition to nonlinear directions. Methods and applications
based on principal curves are developed by Chalmond and Girard
[\citet{chalmondgirard1999}], Dong and McAvoy
[\citet{dongmcavoy1996}], etc. However, the nonlinear principal
curves that are found to explain the most variation in data may not
be interpretable, as is often true for principal components in
linear PCA as well. Izem and Kingsolver [\citet{izemkingsolver2005}]
built a 3-parameter shape invariant
model that decomposes the variation in the data into predetermined
and interpretable directions of interest. Their model describes the
growth rate of families of caterpillars as a function of
temperature:
%
%e1 ###
\begin{equation}
\label{e:SIM}
z_i(t_j)=w_iz\bigl(w_i(t_j-m_i)\bigr)+h_i+\varepsilon_{i,j},
\end{equation}
where $z_i(t_j)$ is the growth rate of family $i$ at temperature $t_j$,
$z$ is the common shape and $h_i$, $m_i$, $w_i$ represent the three
modes of interest: vertical shift, horizontal location and
norm-preserving slope change, respectively. A generalization of (\ref
{e:SIM}) is
%
%e2 ###
\begin{equation}
\label{e:general}
z_i(t_j)=R(\theta_i,t_j)+\varepsilon_{i,j},
\end{equation}
where $z_i(t_j)$ is a nonlinear function of time points $t_j$ and
$\theta_i$ is the vector of parameters that represent fixed modes of variation.

\begin{figure}

\includegraphics{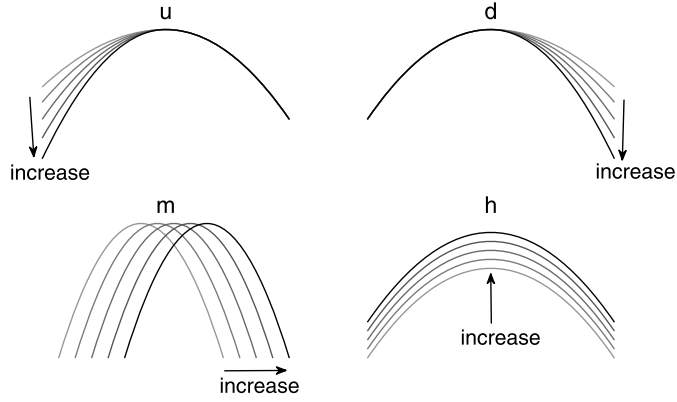}

\caption{Four modes of variation of T-waves:
uphill slope change \textup{(u)}, downhill slope change \textup{(d)}, horizontal location
\textup{(m)} and vertical shift \textup{(h)}.}\label{fig:fourmodes}
\end{figure}

Motivated by their ideas, a model for T-waves is proposed as a
combination of a reference curve and four fixed modes of variation that
are of physiological interest: uphill slope, downhill slope, horizontal
location and vertical shift. The reference curve represents the common
shape of the T-wave; an element in the data matrix $X$ is modeled as
%
%e3 ###
\begin{equation}
\label{e:model} x_{it_j}=x_{i}(t_j)= \cases{
\sqrt{u_id_i}K\bigl(u_i(t_j-m_i)\bigr)+h_i+\varepsilon_{i}(t_j), &\quad
$t_j\leq m_i$,\cr
\sqrt{u_id_i}K\bigl(d_i(t_j-m_i)\bigr)+h_i+\varepsilon_{i}(t_j), &\quad
$t_j> m_i$, }
\end{equation}
where $i=1,\ldots,I$, $j=1,\ldots,J$, $K$ is the reference curve, the
four parameters $u_i$, $d_i$, $m_i$, $h_i$ represent the \textit
{uphill slope}, \textit{downhill slope}, \textit{horizontal
location} and \textit{vertical shift} of the $i$th T wave,
respectively, and $ \varepsilon_{i}(t_j)\sim N(0,\sigma^2_i)$ is the error
term. In light of (\ref{e:general}),
%
%e4 ###
\begin{equation}
\label{e:R}
R(\theta_i,t_j)=\cases{
\sqrt{u_id_i}K\bigl(u_i(t_j-m_i)\bigr)+h_i, &\quad $t_j\leq m_i$,\cr
\sqrt{u_id_i}K\bigl(d_i(t_j-m_i)\bigr)+h_i, &\quad $t_j> m_i$,
}
\end{equation}
where $\theta_i=(u_i,d_i,m_i,h_i)$. Figure \ref{fig:fourmodes} is a
diagram of those four modes of variation.

%f3 ###

The model in (\ref{e:model}) has two innovations over the
3-parameter shape invariant model in (\ref{e:SIM}). First is the
specification of a reference curve. A reference curve represents the
common shape of all curves. It is the curve from which all the other
curves are derived so that the estimated parameters of deformation
can be compared and analyzed. A reference curve differs from the
principal curve since the latter is mathematically defined to
explain the most variation in the data. To measure the variation of
curves within a sample, a natural choice of the reference curve is
the Fr\'{e}chet mean [\citet{frechet1948}] of the curves in this
sample. The Fr\'{e}chet mean is a generalized mean on the nonlinear
manifold, therefore, it represents the common shape of the curves.
However, since the T-waves do not differ greatly in shape and
location, nonparametric methods, such as spline interpolation of the
pointwise average curve to obtain the reference curve, are effective
and much less computationally intensive. (This is illustrated in
Section~\ref{sec3.2}.) The resulting curve is also much closer in shape to the
data than a polynomial-based reference curve as used in
\citet{izemkingsolver2005}.

The concept of reference curve becomes crucial in a multiple sample
analysis, such as a repeated measures design. A repeated measure
design requires a single reference curve, usually a curve obtained
from the baseline sample, so that the changes in the estimated
parameters reflect the changes of the curves over time.

The second
innovation is that this model is a piecewise function of the time
points, since the uphill and downhill curves of the T-wave need to be
modeled separately. This is due to the nature of the T-wave because
the physiologic causes of deformation of the rise and of the decline
of the T-wave can be quite distinct. This method also generalizes to
a model with more piecewise functions that describe distinct shape
changes over different time segments, allowing for more flexibility
in modeling the overall shape.

Using the model in (\ref{e:model}), each T-wave in a given set of
T-waves can be rewritten as a transformation along fixed modes of
variation from a reference curve. Further, such a transformation can be
well approximated by four parameters representing these modes of
variation. This simultaneously achieves significant dimension reduction
of the data and a parametrization with physiological interpretations.

To estimate (\ref{e:model}), minimize the sum of squares of the errors,
that is, for each beat~$i$,
\begin{eqnarray*}
\hat{\theta}_i&=&\operatorname{argmin}\limits_{\theta_i}\sum_j\bigl(x_i(t_j)-R(\theta_i,t_j)\bigr)^2,
\end{eqnarray*}
where $R(\theta_i,t_j)$ is as in (\ref{e:R}). Standard nonlinear
optimization is used to estimate the parameters.

For computational accuracy and efficiency, the reference curves are
centered at the origin both to facilitate the comparison of the
parameter estimates curve to curve and also to minimize interpolation
error. Since the support sets of the reference curve, K, and of X, the
set of observed T-waves, can differ due to changes in the slope
parameters, the support set for K must be extended beyond the support
set of X to allow for interpolation at the endpoints. This causes
interpolation error; centering the reference curve minimizes the change
in the support set and hence reduces this error.

The multiplication factor $\sqrt{u_id_i}$ in (\ref{e:model}) also helps
reduce the change in the support set for the same slope change, and
hence reduces the interpolation error. Note that when $u_i=d_i$, this
factor reduces to $u_i$, which can be regarded as a norm-preserving factor.

It is important to keep in mind that the interpretation of the four
parameters is that they represent the vertical shift, horizontal
location and slope changes of the WHOLE curve, not just any single
point or small part of the curve. They give equal weight to all the
points on the curve, and thus represent the curve better than
measures that are taken from a few points on the curve, such as QT
interval.

%s3.2 ###
\subsection{An illustration}\label{sec3.2}
\begin{figure}[b]

\includegraphics{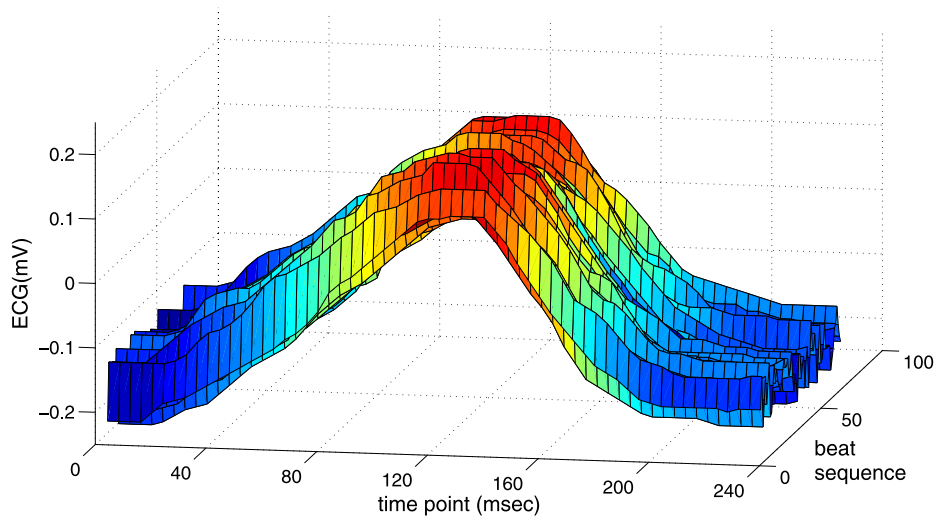}

\caption{Surface plot of a sequence of T-waves of a
sample ECG (the first-minute record of Sel16265).}\label{fig:surf1}
\end{figure}

\begin{figure}

\includegraphics{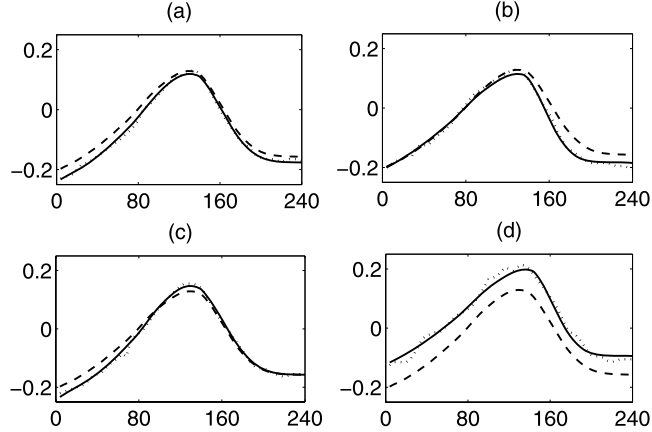}

\caption{Original T-waves (dotted lines), reference
curve (dashed lines) and fitted curves (solid lines) for 4 beats of an
ECG from the QT data base (Sel16265).}\label{fig:fit4}
\end{figure}

A simple example illustrates how the model works. A~one-minute ECG
record from the QT Database (Sel16265) with 66 beats is shown in Figure
\ref{fig:surf1}. The surface plot of the T-waves of this sample
shows the T-waves ``stacked'' in sequence one behind the other. The
color scale, used to accentuate the ``surface geography'' of the 3D
plot, goes from cooler (blues) to warmer (reds) colors over the
range from low to high. Beats are ``stacked'' in sequence one behind
the other. In order to distinguish both the sequence and the color
progression within each single T-wave, the color has been extended
downward as vertical bars of the curve color. Visually, the common
general character of these T-waves is easily described, as is the
variation among them. We apply the model and fit 66 curves based on
the reference curve that is obtained from the average of these 66
curves. Figure \ref{fig:fit4} shows the plots of the T-waves of 4
beats. The dashed lines are the reference curve that is identical
for all beats. The dotted lines are the original T-waves and the
solid lines are the fitted curves. Note that the vertical,
horizontal and slope changes are captured very well by the fitted
curves.
%f4 ###

%s3.3 ###
\subsection{Model robustness to marking of T-wave
boundaries}\label{sec3.3}

\begin{figure}[b]

\includegraphics{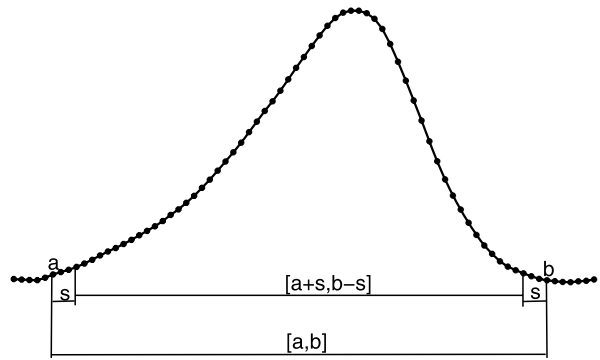}

\caption{Diagram of change in T-wave boundaries
as a function of $s$; $[a,b]$ is the T wave interval by the standard
software. $s=-12,-4,+4,+12$ msec (for ECG at 250 Hz).}\label{fig:stability}
\end{figure}

Accurate determination of the end of the T-wave is widely acknowledged
to be difficult. Therefore, a model that is based on T-wave morphology
rather than T-wave boundaries and that is robust to marking those
boundaries has great potential value. Figure \ref{fig:stability} shows
a T-wave with boundaries $[a, b]$ marked using standard software. Let
\[
[a',b']=[a+s,b-s],
\]
%
%f5 ###
%f6 ###
$s=-12,-4,4,12$ msec. As $s$ takes increasing values from $-$12 msec to
$+$12 msec, the interval changes from the longest one
[$a-12$ msec, $b+12$ msec] to the shortest [$a+12$ msec, $b-12$ msec], and
the model parameters also change. Complete minute records of nine
normal subjects are used for illustration. The robustness of the four
parameter estimates is shown in Table~\ref{table:stabilitynormal}. $\hat{u}_{[a,b]}$ is the
estimate of $u$ at interval $[a,b]$, $\Delta\hat{u}$ is the difference
between $\hat{u}$ at interval $[a+s,b-s]$ and $\hat{u}_{[a,b]}$, that
is, $\Delta\hat{u}=\hat{u}_{[a+s,b-s]}-\hat{u}_{[a,b]}$. The same
definitions apply to $d$, $m$ and $h$. Indicated in the first column,
each measure, median or standard deviation, is adjusted by its scaling
factor, to be comparable across parameters. Notice that both the
medians and the standard deviations are small and are roughly of the
same scale for all the parameters. For parameter $m$, the absolute
change in the unit of milliseconds is also shown.
%
%t1 ###
\begin{table}
\tabcolsep=0pt
\caption{Robustness of $(\hat{u},\hat{d},\hat{m},\hat{h})$ based on
complete minute records for 9 normal subjects}\label{table:stabilitynormal}
\begin{tabular*}{\tablewidth}{@{\extracolsep{4in minus 4in}}lccccc@{}}
\hline\\[-15pt]
\multicolumn{1}{r@{}}{\textbf{Interval:}} & \textbf{Longest} & & & & \textbf{Shortest}\\
& $\bolds{[a-12,b+12]}$ & $\bolds{[a-4,b+4]}$ & $\bolds{[a,b]}$ &$ \bolds{[a+4,b-4]}$ & $\bolds{[a+12,b-12]}$
\\
\hline
median$(\Delta\hat{u})/\hat{u}_{[a,b]}$ & \phantom{$-$}0.0048\phantom{.msec} & \phantom{$-$}0.0024\phantom{.msec} & 0 & $-$0.0004\phantom{.msec}& \phantom{$-$}0.0010\phantom{.msec}
\\
stdev$(\Delta\hat{u})/\hat{u}_{[a,b]}$ & \phantom{$-$}0.0243\phantom{.msec} & \phantom{$-$}0.0096\phantom{.msec} & 0 & \phantom{$-$}0.0065\phantom{.msec} &\phantom{$-$}0.0107\phantom{.msec}
\\[5pt]
median$(\Delta\hat{d})/\hat{d}_{[a,b]}$ & $-$0.0000\phantom{.msec} & \phantom{$-$}0.0009\phantom{.msec} & 0 & \phantom{$-$}0.0006\phantom{.msec}& \phantom{$-$}0.0002\phantom{.msec}
\\
stdev$(\Delta\hat{d})/\hat{d}_{[a,b]}$ & \phantom{$-$}0.0204\phantom{.msec} & \phantom{$-$}0.0079\phantom{.msec} & 0 & \phantom{$-$}0.0055\phantom{.msec} &\phantom{$-$}0.0119\phantom{.msec}
\\[5pt]
median$(\Delta\hat{h})/\hat{h}_{[a,b]}$ & $-$0.0019\phantom{.msec} & $-$0.0023\phantom{.msec} & 0 & $-$0.0011\phantom{.msec} & $-$0.0012\phantom{.msec}
\\
stdev$(\Delta\hat{h})/\hat{h}_{[a,b]}$ & \phantom{$-$}0.0059\phantom{.msec} & \phantom{$-$}0.0063\phantom{.msec} & 0 & \phantom{$-$}0.0055\phantom{.msec} &\phantom{$-$}0.0064\phantom{.msec}
\\[5pt]
median$(\Delta\hat{m})/\hat{m}_{[a,b]}$ & $-$0.0005\phantom{.msec} &$-$0.0000\phantom{.msec} & 0 &$-$0.0000\phantom{.msec} &$-$0.0001\phantom{.msec}
\\
stdev$(\Delta\hat{m})/\hat{m}_{[a,b]}$ & \phantom{$-$}0.0058\phantom{.msec} & \phantom{$-$}0.0014\phantom{.msec} & 0 & \phantom{$-$}0.0012\phantom{.msec} &\phantom{$-$}0.0020\phantom{.msec}
\\
median$(\Delta\hat{m})$ & \phantom{$-$}0.0131 msec & \phantom{$-$}0.0544 msec & 0 & $-$0.0395 msec & $-$0.1684 msec
\\
stdev$(\Delta\hat{m})$ & \phantom{$-$}0.3723 msec &\phantom{$-$}0.1825 msec & 0 & \phantom{$-$}0.2290 msec & \phantom{$-$}0.5124 msec
\\
\hline
\end{tabular*}
\end{table}

A plot provided courtesy of an anonymous referee shows two T-waves,
one following the placebo, the other following administration of a
positive control (amoxicillin). At hour 4, the percentage difference
between uphill slopes for placebo and amoxicillin ($u_p$ and $u_a$,
respectively) is $(u_p-u_a)/u_p$, approximately 0.111. The downhill
slope percentage difference $((d_p-d_a)/d_p)$ is roughly 0.156, the
vertical percentage difference is roughly 0.111, and the horizontal
percentage difference is roughly 0.083. Note that these differences
substantially exceed the magnitude of change in the parameter
estimates under different settings of the T-wave boundaries, as
shown in Table~\ref{table:stabilitynormal}. Thus, the analysis of T-wave morphology is
sufficiently sensitive to detect these drug induced changes. The
median and the standard deviation of $\Delta\hat{m}$, measured in
milliseconds, also show that alteration of the T-wave boundaries do
not present difficulties in the horizontal location estimate, given
that the minimum critical change in the QT (cut-off value) is orders
of magnitude higher, usually 10 milliseconds.

The 24 milliseconds range for $s$ (altering the
interval length over a range of 48 milliseconds) was chosen to
establish robustness for $(u_i,d_i,m_i,h_i)$ over a broad range for
concern about QT prolongation based on FDA practice of seriously
scrutinizing prolongations in excess of 10 milliseconds.

%s3.4 ###
\subsection{Relation of T-wave shape to QT}\label{sec3.4}

\begin{figure}[b]

\includegraphics{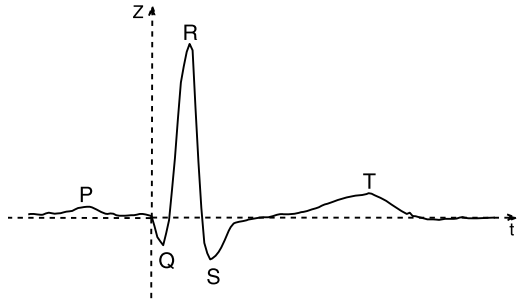}

\caption{To quantify the relationship between QT and
four parameters, the ECG of one beat is placed on a plane where the
baseline matches the horizontal axis $t$, and the start of Q is placed
at the origin. For convenience, QT is measured from the origin to the
point where the T-wave intersects the horizontal axis on its right.}\label{fig:ECGQT}
\end{figure}

Effectively, the relationship between the four parameters that describe
changes in the shape of the T-wave and changes in the QT interval is
through translation and/or through flattening of the T-wave. As shown
in Figure \ref{fig:ECGQT}, measurement of QT can be depicted by placing
one beat on a plane where the baseline matches the horizontal axis,
$t$, and with the origin placed at the initiation of the Q-wave. The QT
interval then ends approximately where the T-wave intersects the
horizontal axis on its right. To see the relationship between the four
parameters and QT, first consider their relationship to be a (less
complicated) quadratic form:
%
%e5 ###
\begin{equation}
\label{e:quadratic}
K(t)=a(t-b)^2+c,\qquad a<0, c>0;
\end{equation}
QT can be calculated explicitly:
%
%e6 ###
\begin{equation}
\label{e:QT}
QT=b+\sqrt{-\frac{c}{a}}.
\end{equation}

Let $\tilde{K}(t)$ be the quadratic function written in the form of the
model in (\ref{e:model}),
\begin{eqnarray*}
\tilde{K}(t)&=&\cases{
\sqrt{u_id_i}\bigl[a\bigl(u_i(t-m_i)-b\bigr)^2+c\bigr]+h_i, &\quad $t\leq m_i$,\cr
\sqrt{u_id_i}\bigl[a\bigl(d_i(t-m_i)-b\bigr)^2+c\bigr]+h_i, &\quad $t> m_i$.
}
\end{eqnarray*}
Thus,
%
%e7 ###
\begin{equation}
\label{e:QTnew}
\tilde{QT}=\cases{
m_i+\dfrac{b}{u_i}+\dfrac{1}{u_i}\sqrt{-\dfrac{c}{a}-\dfrac{h}{a\sqrt
{u_id_i}}}, &\quad $t\leq m_i$,\cr
m_i+\dfrac{b}{d_i}+\dfrac{1}{d_i}\sqrt{-\dfrac{c}{a}-\dfrac{h}{a\sqrt
{u_id_i}}}, &\quad $t> m_i$.
}
\end{equation}
%
%f7 ###
From (\ref{e:QTnew}) the dependence of QT on $m$ is seen to be direct
and positive, while dependence on $u$ and on $d$ is inverse. The
relationship of $(h)^{1/2}$ to QT is direct, but also involves both $u$
and $d$. Therefore, when, as can occur in practice, the data show $\hat
{h}$ to be correlated with $\hat{u}$ and/or $\hat{d}$, the observed QT
may or may not exhibit a positive simple correlation of QT with $\hat{h}$.

%f8 ###
\begin{figure}[b]

\includegraphics{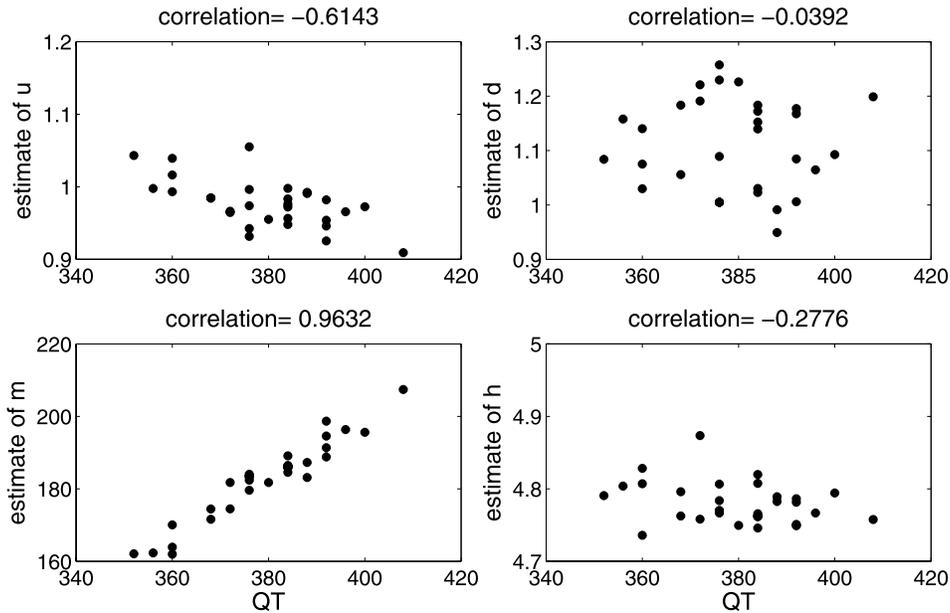}

\caption{Four parameter estimates versus QT
(Sel103) showing correlations with $\hat{m}$ (positive) and $\hat{u}$
(negative).}\label{fig:QTwwmh103}
\end{figure}

One example that serves as an illustration is a half-minute record
(Sel103), an arrhythmia from the QT Dataset (with QT calculated using
ECGpuwave software also available on \href{http://www.physionet.org}{www.physionet.org}). Figure \ref
{fig:QTwwmh103} shows the scatterplots of the parameter estimates and
QT intervals. For this record, QT dependence on $\hat{m}$ is evidenced
in their positive correlation ($r=0.9632$), indicating that the T-wave
shifts to the right to increase QT. The negative correlation between
$\hat{u}$ and QT means that as the uphill curve flattens, QT gets
longer. This confirms expert qualitative statements about T-wave changes.

%s4 ###
\section{Statistical inference}\label{sec4}

The principal objectives for a functional data analysis approach to
analyze sequential T-waves are as follows: (1) to use information for
the complete form of a T-wave, (2) to capture full information for an
extended series of beats, and (3) to define a low-dimensional
parametrization model to use in drawing statistical inferences.

Consider the problem arising in analysis of an arrhythmia:
classification of beats according to the shape of their T-waves.
Clustering similar beats in ECGs is the first step in identifying
patterns of beats that characterize specific cardiac function
abnormalities. For example, if a cluster of similar beats is defined in
terms of the parameter $h$ (and not dependent upon $u$, $d$ or $m$),
then the types of beats differ by the T-wave heights, that is, by the
signal intensity not by its pattern. In contrast, multivariate-defined
clusters differ according to the signal patterns and possibly the
signal intensity.

%f9 ###
\begin{figure}[t]

\includegraphics{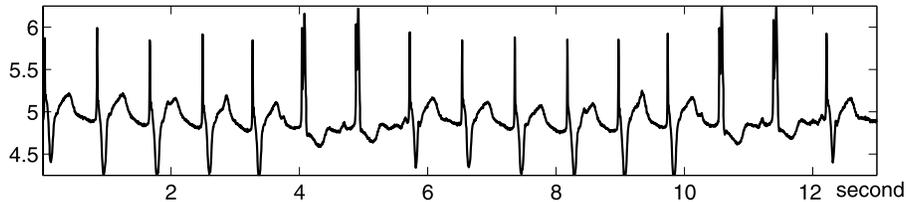}

\caption{A sample ECG for patient Sel104.}\label{fig:ECGtracing}
\end{figure}
%f10 ###
\begin{figure}[b]

\includegraphics{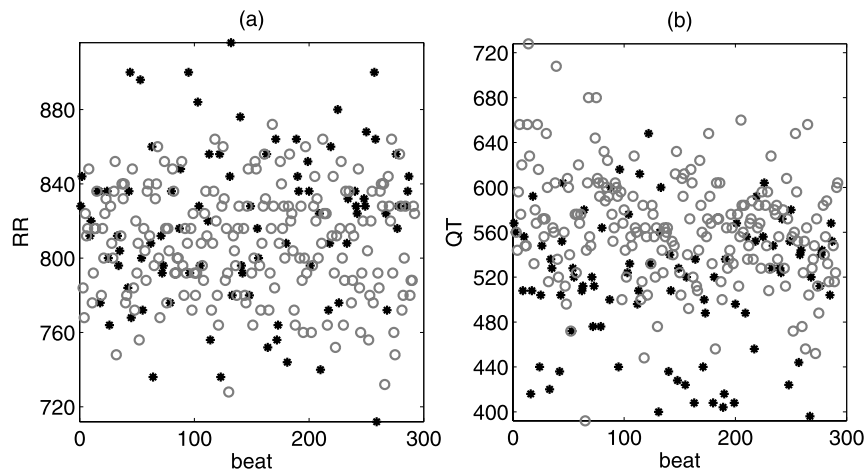}

\caption{Normal (circle) and abnormal (star)
beats, plotted vs \textup{(a)} RR and \textup{(b)} QT.}\label{fig:class104RRQT}
\end{figure}

Figure \ref{fig:ECGtracing} shows a sequence of the ECG tracing of
an arrhythmia subject Sel104 from the QT Dataset (a four-minute record
with 292 beats). Visually there are two major types of beats: one
type with normal T-waves and the other type with ``S'' shape T-waves.
Figure \ref{fig:class104RRQT} shows that neither the beat length
(RR: interval from preceding R peak to succeeding one) nor QT
successfully discriminates between these two types of beats. (Here
the QT intervals are obtained by applying ECGpuwave software
followed by confirmation or correction by expert review.)

\begin{figure}

\includegraphics{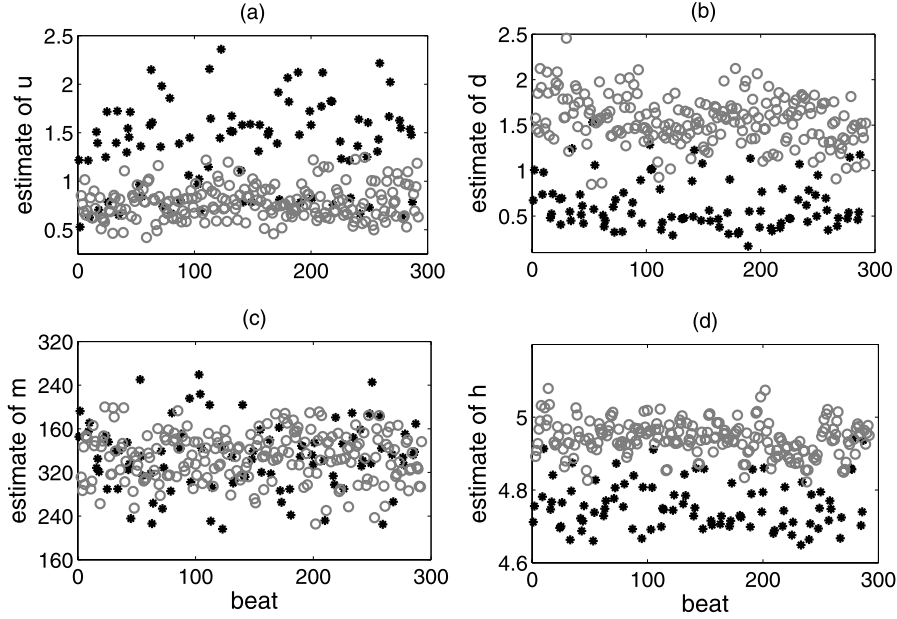}

\caption{Normal (circle) and abnormal (star)
beats for a 4-minute record (292 beats) of Sel104, plotted as time
series of \textup{(a)} $\hat{u}$ \textup{(b)} $\hat{d}$ \textup{(c)} $\hat{m}$
and \textup{(d)} $\hat{h}$.}\label{fig:class104udmh}
\end{figure}

Figure \ref{fig:class104udmh} shows the plots of $\hat{u}$, $\hat{d}$,
$\hat{m}$ and $\hat{h}$ in (a), (b), (c) and (d), respectively. Observe
that both $\hat{d}$ and $\hat{h}$ do well in separating these two
groups. In fact, using K-means clustering for $\hat{d}$, one can get
two clusters that match (95.35\% of beats) with the two groups. $\hat
{u}$ also does a fairly good job, but $\hat{m}$ does not distinguish
the two groups.

%f11 ###

An alternative method for modeling T-wave shape uses two Gaussians.
Approximation of a single T-wave as a mixture of two Gaussians can
be quite precise [\citet{clifford2006}]; this can be combined with a
suitable algorithm to define the end of T-wave in terms of a
specified tail probability. However, this exercise is (independent)
curve fitting to individual beats, hence is not amenable to further
statistical inference.

\begin{figure}

\includegraphics{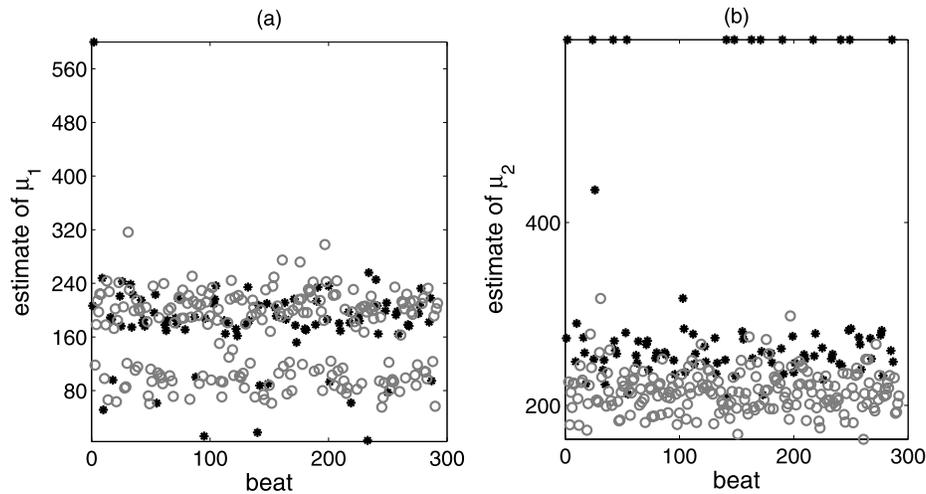}

\caption{Normal (circle) and abnormal (star)
beats, plotted vs \textup{(a)} the first location parameter $\mu_1$ of a
Gaussian model and \textup{(b)} the second location parameter $\mu_2$ of a
Gaussian model.}\label{fig:class104Gau}
\end{figure}

\begin{figure}

\includegraphics{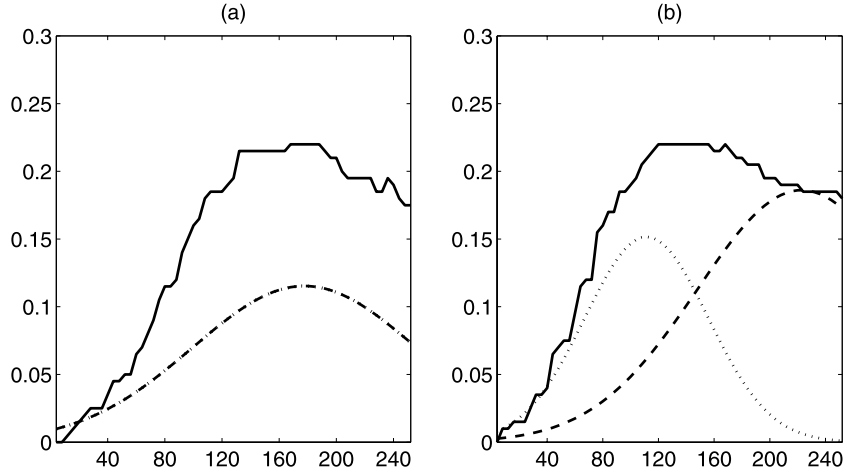}

\caption{Two similar T-waves (solid lines) in record
Sel104, with two Gaussian functions (dashed lines and dotted lines)
fitted to each by the Gaussian model.}\label{fig:Gau104}
\end{figure}

Figure \ref{fig:class104Gau} plots the means $(\mu_1,\mu_2)$ of the
left and right Gaussians fitted as a mixture. The left Gaussian which
dominates the onset of the T-wave does not discriminate between beat
types. The right Gaussian which dominates extreme tail behavior is only
partially successful, with an overall misclassification rate of $26.73\%
$. (By comparison, the overall misclassification rate using the
four-parameter model is $4.65\%$.) However, inferences about the
typical shape of each type of T-wave remains difficult. This is due in
part to the number of estimated parameters (6 for each beat) and in
part to the instability of the parametrization. In fact,
parametrization of the Gaussian model is not robust to small changes in
T-wave shape, such as those caused by noise. As an illustration, Figure
\ref{fig:Gau104} shows 2 T-waves in record Sel104. The very slight
difference in the last part of the two T-waves induces two very
different parametrizations. For the first T-wave in (a), the two
Gaussian functions are almost identical. The parametrization $(\lambda
_i, \sigma_i^2, \mu_i)$, where $\lambda_i$ is the unnormalized mixing
coefficient for the $i$th Gaussian, is (4.5102, 19.5607, 44.4245) for
the first Gaussian and (4.5102, 19.5588, 44.4265) for the second
Gaussian; and the two Gaussian curves are seen to overlap in Figure \ref
{fig:Gau104}(a), while for the second T-wave in (b) the two sets of
parameters are quite different: (3.3558, 11.0680, 27.8644) and (6.9715,
18.7290, 55.7295), and the Gaussians separate as shown in Figure \ref
{fig:Gau104}(b). Thus, the parameters for the Gaussian model cannot
accurately reflect the degree of difference among curves, and hence do
not make a good biomarker for analysis of T-wave shape or statistical
inference about beat problems.

%f12 ###

%f13 ###

% 4.5102 4.5102 19.5607 19.5588 44.4245 44.4265
% 3.3558 6.9715 11.0680 18.7290 27.8644 55.7295

\begin{figure}[b]

\includegraphics{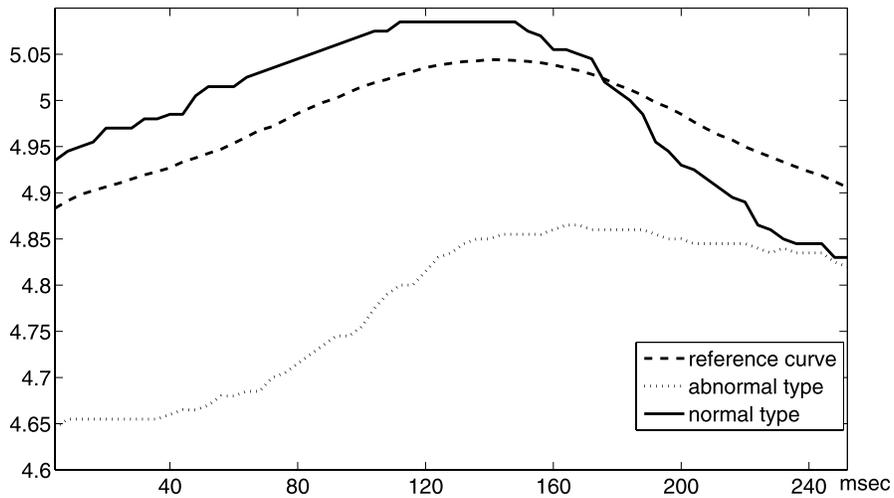}

\caption{The reference curve and typical curves
of the two types of T-waves in Sel104.}\label{fig:Tshape104}
\end{figure}

There are other methods that work well in classifying curves, such
as wavelet-based methods [\citet{wangraymallick2007}], or even PCA;
but neither wavelet coefficients nor principal components of the
curves help to understand the physiological change in the shape of
the T-wave. From the sequences of four parameter estimates one can
obtain more information about the different shape of the T-wave for
different groups. For example, by inspecting Figure
\ref{fig:class104udmh}, one observes that, among the two groups, one
group has higher $\hat{u}$ (greater than 1), lower $\hat{d}$ (less
than 1), lower $\hat{h}$ and similar $\hat{m}$ as the other. Note
that for $\hat{u}$ and $\hat{d}$, 1 is a cut-off value because when
$\hat{u}>1$, the uphill curve is steeper than the reference curve,
elsewhere it is flatter. The same applies to $\hat{d}$. So it can
be imagined that one group of beats has steeper uphill curve,
flatter downhill curve and lower height than the reference curve.
The other group behaves conversely. The horizontal positions do not
distinguish the two. These descriptions match the true curves as
shown in Figure \ref{fig:Tshape104}.

%f14 ###

\begin{figure}

\includegraphics{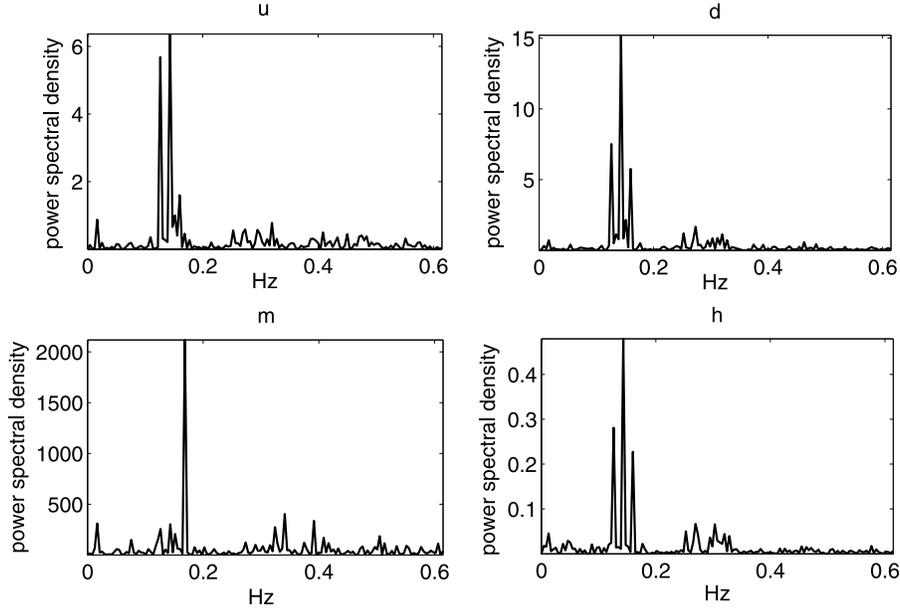}

\caption{The power spectral density of $\hat{u}$,
$\hat{d}$, $\hat{m}$ and $\hat{h}$.}\label{fig:psd104}
\end{figure}

\begin{figure}

\includegraphics{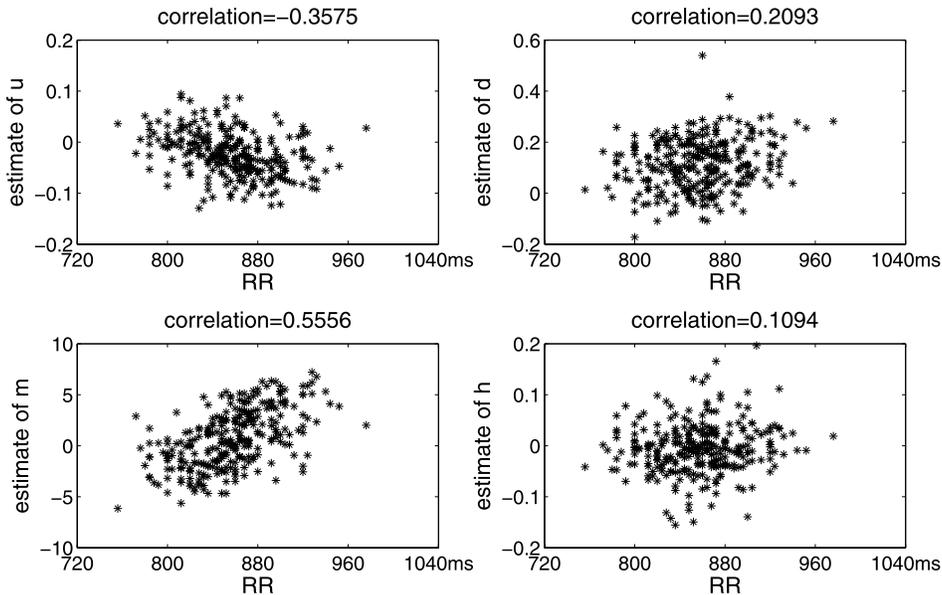}

\caption{Correlations of four parameter
estimates with RR (Sel103).}\label{fig:rrwwmh103}
\end{figure}

Furthermore, one can obtain information about T-wave shape change by
studying the parameter estimates over beats as four time series. To
study the frequency components of these time series, one can obtain the
power spectral density, shown in Figure \ref{fig:psd104}. Note that
$\hat{u}$, $\hat{d}$ and $\hat{h}$ all have peaks at 0.1265 Hz, 0.1429
Hz and 0.1592 Hz, corresponding to periods 7.9 sec (9.6 beats), 7 sec (8.51
beats) and 6.28 sec (7.64 beats). These are roughly the frequencies of
the abnormal beats that can be observed from the ECG chart. $\hat{m}$
has a peak at 0.1674 Hz, corresponding to 5.97 sec (7.2 beats). Since
$\hat{m}$ is most correlated with RR among the four parameters (see
Figure \ref{fig:rrwwmh103}), and regular changes in RR are usually
related to breathing, a reasonable guess is that this frequency
reflects the breathing pattern of this subject. Many other properties
in the time domain and frequency domain can be studied as well.

%f15 ###

In order to make comparisons between time segments or between
experimental conditions, the QT measure needs to be adjusted by the
RR (the inverse of the heart rate) because of the relationship
long-recognized between the two. Although the literature on choices
of adjustment function is extensive, no consensus has been reached
on the optimal ``correction'' or adjustment function. This may be at
least partially attributable to the disparity in changes in the QT
and in the TQ intervals with change in RR, especially in normal
subjects. Adjustment methods for the four-parameter model will
depend on the direct relationship of the T-wave shape and RR; and
this relationship may differ between normal subjects or among
subjects with different known arrhythmias. Also, the dependence
between RR and the onset of T-wave shape alteration may exhibit a
lag of several beats.

The first example, shown in Figure \ref{fig:rrwwmh103}, is a
four-minute record of an arrhythmia subject Sel103 from the QT Dataset.
Here $\hat{u}$ and RR are modestly negatively correlated, which
means that as uphill curve gets flatter, RR gets longer. The
positive correlation of $\hat{m}$ and RR means that the RR
prolongation is generally related to a T-wave shift to the right.
Relationships of $\hat{d}$ and $\hat{h}$ with RR are not apparent in
this record. This data set benefits from a multivariate clustering
approach since several individual relationships between model
parameter estimates and RR are accompanied by interdependencies
among the parameters. Further analysis shows that $\hat{u}$ and
$\hat{d}$ are negatively correlated in this case, that is, the uphill
curve and downhill curve change their slopes in such a way as to
keep the angle between them relatively stable.

%f16 ###

Different arrhythmias exhibit different patterns and may arise from
different causes; the four-parameter model enables inferences about
these patterns. A second example illustrates a different phenomenon. In
this arrhythmia, the sequence of beats can be shown to have an
approximate periodicity. By representing this record as a series of
four-dimensional vectors $(\hat{u},\hat{d},\hat{m},\hat{h})'$, the lag
of principal time-dependency can be established. Table~\ref{table:rrwwmh123} shows the
correlation of each parameter at time $t$ with RR at time $t$, $t-1$,
$t-2$ and $ t-3$ for the 2nd minute of the record of Sel123 from the QT
Dataset. Note that all the parameters are most strongly correlated with
RR at $(t-2)$, indicating a two-beat lag between RR and the altered
T-wave. Thus, a longer beat is followed by alteration in the T-wave
form (shallower). Physiologically this may have explanation in terms of
energy expenditure and the adequacy of the ``rest'' between the end of
the T-wave and the onset of the subsequent P-wave. The last row
indicates that QT does not capture such a phenomenon.

%s5 ###
\section{Discussion}\label{sec5}
Besides the applications mentioned in Section \ref{sec4}, there are others
such as outlier beat detection and discrimination between normal
subjects and arrhythmia subjects. One can detect outlier beats by
treating the four parameters' values as four-dimensional vector data
and applying standard outlier detection methods for multivariate
data. Distinguishing arrhythmias from normal heart rhythms can be
done based on the patterns of variation in this four-dimensional
characterization. Work is ongoing to develop specific methodology to
encompass beat classification and analysis of the temporal process.

Methods for analyzing data with a general data structure described
in Section~\ref{sec2.2} depend on the purpose of the study. For the
measurement of T-wave variation within each sample, for example, detection
of outlier beats within a record, the reference curve can be
computed as the pointwise average curve for the record, and the
analysis is based on the parameter estimates for the curves in that
record. For multiple records of the same subject, a
``super-reference curve'' can be computed, by analogy to MANOVA
methodology, the parameter estimates using the within-record
reference curve and those computed using the super-reference curve
can be analyzed. (This approach inherits the same problems of
unequal variances and unequal sample sizes that are present in
MANOVA analyses with essentially the same solutions.)

%t2 ###
\begin{table}
\tablewidth=200pt
\caption{Correlation between the estimated model parameters and current
and previous $RR$s}\label{table:rrwwmh123}
\begin{tabular*}{200pt}{@{\extracolsep{4in minus 4in}}lcccc@{}}
\hline
\textbf{RR} &$\bolds{t}$ &$\bolds{t-1}$ &$\bolds{t-2}$ &$\bolds{t-3}$ \\
\hline
$\hat{u}$ & $-$0.3216 &$-$0.4003 &$\bolds{-}$\textbf{0.7554} &$-$0.3152 \\
$\hat{d}$ & $-$0.4580 & $-$0.6309 &$\bolds{-}$\textbf{0.7428} &$-$0.3957 \\
$\hat{m}$ & \phantom{$-$}0.3009 &\phantom{$-$}0.3419 &\phantom{$-$}\textbf{0.4596} &\phantom{$-$}0.1900 \\
$\hat{h}$ & \phantom{$-$}0.2410 &$-$0.1825 &$\bolds{-}$\textbf{0.3977} &\phantom{$-$}0.0605 \\
QT & $-$0.0478 & \phantom{$-$}0.1953 & \phantom{$-$}0.1646 & $-$0.0455 \\
\hline
\end{tabular*}
\end{table}

For complex experiment designs with ECGs taken repeatedly over time
and/or under varying conditions, the actual time courses of T-wave
changes as well as variation may be the primary focus. (For
example, in a study of a new drug, ECGs may be conducted for each
subject at a dozen time points each day beginning with a baseline---none,
placebo, positive control substance, drug, etc.) In such a
case the reference curves for individual records within each set
form a data set of curves to be studied. Once again, applying a
functional analysis approach, a hyper-reference curve can be
computed (from the record reference curves) and analyzed.

The functional data approach used here could also be extended to
alternative choices of the reference curve. For example, robust
methods could be applied to reduce the influence of outlier beats.
In the case when there are clusters of beats varying by shape, such
as in Figure~\ref{fig:ECGtracing}, this approach may first be applied to cluster beats,
as described in Section~\ref{sec4}, then it is applied within each cluster to
capture the individual differences within the cluster.

Extension of the method proposed here to multi-lead ECG data is
direct. Standard 12- or 16-lead ECGs provide spatial information
about the heart as well as redundancies and ``check information.''
Current work combines dimension reduction methods with this
functional data modeling approach to more fully incorporate the
information from multiple leads without decreasing the
signal-to-noise ratio. In addition, since specific leads measure
the electrical potential across different parts of the heart, the
extension to a higher dimensional model should lead to a more
general methodology and increase sensitivity to other aberrant
cardiac behaviors.

%s6 ###
\section{Conclusion}\label{sec6}

In this paper functional data analysis is used to construct a
statistical model of ECG T-waves. This model makes the physiologically
reasonable assumption that there is a common primary shape (within
subject) for the T-waves and it uses four interpretable parameters to
describe the individual deviation of each beat from the common T-wave
shape. The model accounts for the entire T-wave morphology, making the
estimation robust to the marking of T-wave boundaries. Applications
such as classification of beats were illustrated for this model.
Application of this model to measuring drug-induced change in T-wave is
intended, pending availability of control ECGs from actual QT studies.

Note: Programs in matlab to implement the method are available from the
authors. Please send requests by email to \ead[label=e3]{yingchun\_z@yahoo.com}\printead*{e3}.

\section*{Acknowledgments}
We thank the NISS Pharmaceutical Affiliates for\break proposing this area of
research and for providing continuing assistance. We also thank
Professor George Moody of MIT for introducing us to physionet as a data
source; and we thank Dr. Stan Young of NISS for helpful conversations
on this topic.

\printaddresses

\end{document}